\documentclass{iau}
\usepackage{graphicx}

\title[MWC\,656: A Be+BH or a Be+sdO] 
{MWC\,656: A Be+BH or a Be+sdO?}

\author[Rivinius et al.]   
{Th.~Rivinius$^1$, R. Klement$^{2}$, S.~D. Chojnowski$^{3}$, D. Baade$^{4}$, K.~Shepard$^{5}$ \and P. Hadrava$^{6}$\\
}

\affiliation{$^1$ESO - European Organisation for Astronomical Research in the Southern Hemisphere, Chile\\ email: {\tt triviniu@eso.org} \\[\affilskip]
  $^{2}$The CHARA Array of Georgia State University, Mount Wilson, USA\\[\affilskip]
  $^{3}$Department of Physics, Montana State University, USA \\[\affilskip]
  $^{4}$ESO - European Organisation for Astron.\ Research in the Southern Hemisphere, Germany\\[\affilskip]
  $^{5}$ CHARA and Department of Physics and Astronomy, Georgia State University, USA \\[\affilskip]
  $^{6}$ Astronomical Institute of the Academy of Sciences of the Czech Republic, Czech Republic
}
\pubyear{2022}
\volume{361}  
\pagerange{}
\jname{Massive Stars Near and Far}
\begin{document}

\maketitle

\begin{abstract}MWC\,656 has been reported as classical Be star with a black hole companion. Revisited spectral variability properties render this unlikely, with a hot subdwarf more probable.
\end{abstract}

\firstsection 
\section{Introduction}

\noindent 
\cite{2014Natur.505..378C} identified MWC\,656 (HD\,215227) as a binary consisting of a classical Be star and a black hole (BH). The interest in this star was triggered by a $\mathbf\gamma$-ray detection with 0.6$^{\circ}$ error circle (\cite[Williams et al. 2010]{2010ApJ...723L..93W}). 
The identification of a BH in MWC\,656 hinges on radial velocities (RVs) from optical spectra, most critically on {Fe}{\sc ii}\,$\lambda$4583 emission from a circumstellar disk tracing the motion of the Be primary.  

More recently, however, Be stars are being recognized in large numbers as Be+sdOB systems, i.e., with hot subdwarf companions (\cite[Wang et al. 2021]{2021AJ....161..248W}), some showing striking similarities to MWC\,656.  
For a  conclusive distinction between a Be+BH and a Be+sdO system, we re-examined the original data together with new higher-resolution spectra from the past decade, namely the original 32 blue and 64 red spectra of \cite[Casares et al.]{2014Natur.505..378C} (R=5500), four echelle spectra from amateur observers (R=9\,000-12\,000) from the BeSS database, three epochs of linear polarimetry, i.e., with very high $S/N$, from ESPaDOnS (R=48\,000) at the CFHT, and
twelve spectra from ARCES (R=31\,500) at Apache Point Observatory.
%

%
%
\section{Results}

\noindent {\bf Re-determination of the orbital period:}
The combined spectroscopic database proved incompatible with the period of $60.37\pm0.04$\,d found by \cite{2010ApJ...723L..93W} in suboptimal photometric data and adopted by \cite{2014Natur.505..378C}.  Fourier analysis of our new RV measurements from fitting a Gaussian to all the relatively narrow {He}{\sc ii}\,$\lambda$4686 emission profiles suggests a similar and highly significant period of $59.12\pm0.05$\,d.

\noindent {\bf Radial velocity amplitudes:}
The semi-amplitude, $K_2$, of {He}{\sc ii}\,$\lambda$4686 is virtually identical to that obtained by \cite{2014Natur.505..378C} from the same line, i.e., about 80\,km/s (Fig.~1, left).
However, the optically thin double-peaked emission lines from the disk surrounding the Be star \textit{do not} simply shift due to orbital motion, but also vary in peak separation (Fig.~1, middle).  This is due to the tidal distortion of the disk (\cite[Panoglou et al. 2016]{2016MNRAS.461.2616P}) and makes the line peaks unsuitable for measuring the RV amplitude of the Be star, ${K_1}$.  The effect is well illustrated also by the base of H$\alpha$\ that changes in width with the orbital period.  Since \cite{2014Natur.505..378C} used such lines to measure $K_1$, this invalidates their results and thus also the mass ratio which led to the suggestion of a BH.  

In the new spectra, we detected photospheric lines from the Be star, in particular {He}{\sc i}\,$\lambda$6678 (Fig.~1, right).  The outer edges have peak-to-peak RV amplitudes of about 20\,km/s\ (red side) and 30\,km/s\ (blue side), respectively. 
The implied amplitude of $K_1$ of 10 to 15\,km/s is at least a factor of 2 less than the value of $32$\,km/s derived by \cite{2014Natur.505..378C}.  At $K_1=7$\,km/s, the RV amplitude of H$\alpha$\ (measured via the bisector of the lower wings) is even lower.  This correction reduces the mass of the companion to the Be star by a similar factor and probably removes it from the BH domain.  

\section{Discussion}

\noindent
The {He}{\sc ii}\,$\lambda$4686 emission, from which the RV amplitude, $K_2$, of the companion to the Be star was determined, is double-peaked and arises from an accretion disk around the companion.  It is fed by spillover from the disk around the Be star.  Such variable emission is also seen in the well-known Be+sdO binaries $\varphi$\,Per, 59\,Cyg, and HD\,55606 (\cite[Chojnowski et al. 2018]{2018ApJ...865...76C}). 
In addition, there are circumstellar gas streams, also similar to many sdO stars: {Fig.~1} shows the new spectra phased with the orbital period:  The  {He}{\sc i}\,$\lambda$6678 emission traces the orbit of the secondary, just as {He}{\sc ii}\,$\lambda$4686 does.  But there is also an additional, phase locked \textit{absorption} component. Around phase 0.4 it grows into a very deep shell absorption lasting for about 15\% of the cycle.
Owing to the lack/weakness of X-rays, \cite{2014Natur.505..378C} posited that the BH is quiescent, i.e., non-accreting.  The improved spectroscopic mapping of the gas in the system contradicts this.

Hence, MWC\,656 is more likely to host a hot-subdwarf instead of a black-hole, and its proximity no longer serves to indicate a relatively large number of Be+BH binaries.


\begin{figure}[b]
\begin{center}
\includegraphics[width=0.33\textwidth,clip,viewport=17 52 450 563]{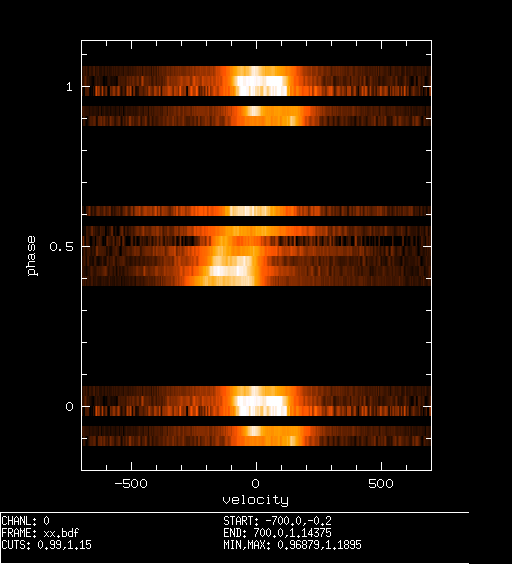}%
\includegraphics[width=0.33\textwidth,clip,viewport=17 52 450 563]{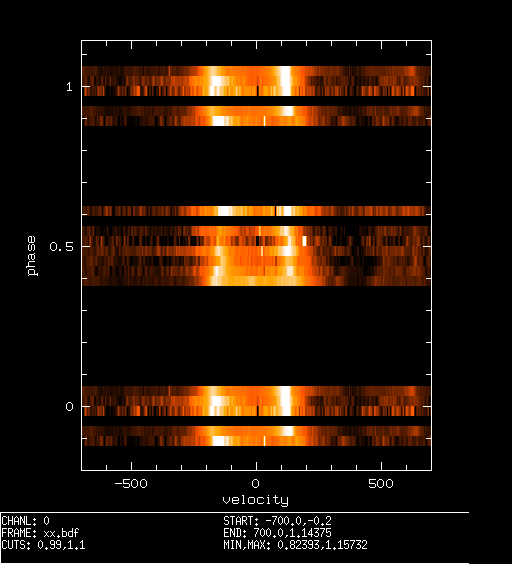}%
\includegraphics[width=0.33\textwidth,clip,viewport=17 52 450 563]{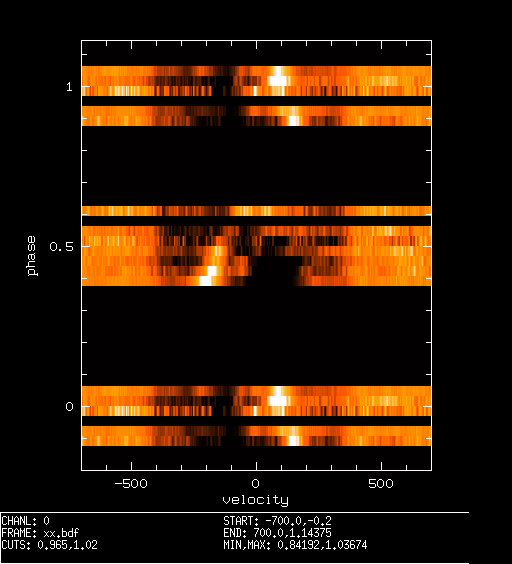}%
 \caption{New spectra folded with revised period. $\pm$20\% phase extensions are shown for clarity. From left to right, the spectral lines are {He}{\sc ii}\,$\lambda$4686, {Fe}{\sc ii}\,$\lambda$5317, and {He}{\sc i}\,$\lambda$6678.  As phase zero, we adopt the periastron time given by \cite{2014Natur.505..378C}.}
\end{center}
\end{figure}

\end{document}